\documentclass[11pt]{article}
\usepackage{amssymb}
\usepackage{latexsym}
\usepackage[dvips]{graphicx}

\newcommand{\be}{\begin{equation}}
\newcommand{\ee}{\end{equation}}
\newcommand{\beaa}{\begin{eqnarray}}
\newcommand{\eeaa}{\end{eqnarray}}
\newcommand{\sacolsep}{\setlength\arraycolsep{2pt}}
\newcommand{\ds}{\displaystyle}

\newcommand{\ovp}{\overline{\Phi}}
\newcommand{\cb}{\bar{\chi}}
\newcommand{\pb}{\bar{\psi}}
\newcommand{\etab}{\bar{\eta}}
\newcommand{\phipv}{\Phi^{\textrm{\tiny{PV}}}}
\newcommand{\jpv}{\mathcal{J}^{\textrm{\tiny{PV}}}}
\newcommand{\jtpv}{\tilde{\mathcal{J}}^{\textrm{\tiny{PV}}}}
\newcommand{\dslash}{\not{\hspace{-0.3 em}D}}

\newcommand{\Qslash}{\not{\hspace{-0.25 em}Q}}
\newcommand{\iphi}{\int \mathcal{D}\Phi \;}
\newcommand{\ipsi}{\int \mathcal{D}\Psi \;}

\newcommand{\dpdpb}{\mathcal{D}\psi \mathcal{D}\pb \;}

\newcommand{\Tr}{\mathrm{Tr}}
\newcommand{\Trs}{\mathrm{Tr}\,}
\newcommand{\Trg}{\Tr_{\textrm{\tiny{G}}}}
\renewcommand{\l}{\Lambda}

\newcommand{\M}{\mathcal{M}}
\newcommand{\tinyb}{\textrm{\tiny{B}}}

\author{S. Chiantese \\ \emph{\small{Dipartimento di
Fisica,}} \\ \emph{\small{Universit\`a degli Studi di Roma
``La Sapienza''}} \\ \emph{\small{P. le Aldo Moro, 2 - 00185
Roma, Italia}} \\ \emph{\small{E-mail:
stefano.chiantese@roma1.infn.it}}}
\title{\textbf{Non-diagrammatic calculation of QCD \\ one-loop}
$\mathbf{\beta}$\textbf{-function based on the \\
renormalization group equation}}
\date{}
\begin{document}
\maketitle
\begin{abstract}
Using the higher covariant derivatives regularization of
gauge theories in the framework of the background field
method, supplemented with one-loop Pauli-Villars regulator
fields, we obtain a version of the renormalization group
equation for the regulator fields, whose vacuum energy
depends on the background gauge field. It is evaluated using
an anomalous Ward-Takahashi identity, which is related to
the rescaling anomaly of the auxiliary fields, obtained by
the Fujikawa approach. In this way the anomalous origin of
the one-loop $\beta$-function in QCD is clearly shown in
terms of scaling of effective Lagrangians without the use of
any Feynman diagram. The simplicity of the method is due to
the preservation of the background and quantum gauge
invariance in any step of the calculation.
\end{abstract}

\section*{Introduction}

In an interesting paper Polchinski demonstrated that
Wilson's decimation method \cite{wilson} applied to
continuum field theory is sufficient to provide proof of
perturbative renormalization \cite{polchinski}. To obtain
the exact renormalization group equation (ERGE) he used an
obvious identity, which consists in setting the integral
functional of a proper total derivative to zero, and the
independence of the partition function on the scale. Such an
idea corresponds to a reparameterization of the partition
function since the total derivative emerges from a field
redefinition \cite{morris0, latorre}.

This derivation of ERGE has to be modified when external
gauge fields are present. In such a case, as they could be
field dependent, we can not discard the singular terms that
appear when developing the total derivative before having
studied their possible physical relevance. In fact, as was
shown in ref. \cite{yoshida} in the context of $N=1$
supersymmetric Yang-Mills theory, they assume the meaning of
a vacuum energy and are responsible for the exact one-loop
running of the holomorphic gauge coupling \cite{shifman}.

Here the analysis in \cite{yoshida, yoshidac} is extended to
non-supersymmetric gauge theories. In those papers the gauge
invariant regularization proposed by Arkani-Hamed and
Murayama \cite{murayama} is used, which consists in giving a
big mass to the extra fields\footnote{With extra field we
mean a field of the finite theory which does not appear in
the theory we are regulating.} of a finite theory with
extended supersymmetry. In conventional gauge theories we
clearly have to resort to a different regularization. The
attractive properties of gauge invariance, non-perturbative
meaning and applicability to chiral and supersymmetric
models make the regularization proposed by Slavnov
\cite{slavnov1} interesting. It is a hybrid of higher
covariant derivatives and Pauli-Villars (PV)
regularizations. Nevertheless it has some inconsistencies,
the main one being known as overlapping divergences.
Although minor modifications of the original scheme are
possible, which make the regularization self-consistent
\cite{falceto, slavnov2}, it is not yet known how to use it
in the RG context.

In this paper a solution to this is offered at the one-loop
level using the background field method when the regulator
fields are the only to flow. The outcome is a version of RG
equation close to the ones in \cite{yoshida, yoshidac}, from
which we obtain the one-loop $\beta$-function of
non-supersymmetric $SU(N)$ Yang-Mills theory without using
Feynman diagrams. The simplicity of the calculation is a
result of the preservation of the background and quantum
gauge invariance.

The background field method is technically useful for
calculating the vacuum energy of regulator fields, but it
also introduces conceptual simplifications. For instance,
using the regularization mentioned above in its framework, a
formal transition to a covariant background gauge is not
required to prove the gauge invariance of one-loop
divergences. The invariance of the partition function under
gauge transformations of the background field makes it
evident.

The paper is structured as follows. Noting that the gauge
field can be considered external for the calculation of the
one-loop $\beta$-function in QED, we begin with the Abelian
theory to show how our method works without taking into
account complications which are due to the quantum
fluctuations. In section \ref{secgiea}, recording the
regularized gauge invariant effective action of the
non-Abelian theory, we emphasize the important points of the
Slavnov regularization and the background field method for
our approach. In section \ref{secbna} the calculation of the
one-loop $\beta$-function is performed when the gauge group
is $SU(N)$. The conclusions are followed by two appendices,
the first one reporting the derivation of an equation that
we shall term 't Hooft's and the second one the calculation
of the Jacobians used in the text.

\section{One-loop $\beta$-function of QED}
\label{secba}

As a result of Ward's identity in QED, it is a well known
fact that the charge renormalization originates solely from
vacuum polarization. Then, at the one-loop level, the
quantum fluctuations of the gauge field can be disregarded
to achieve the $\beta$-function\footnote{Incidentally, with
the gauge field treated as external, the vacuum polarization
is the only divergent diagram. In fact, indicating with $N$
the number of vertices, the spinor cycles are divergent when
$N<5$, $N=2$ being the maximum grade. The cycle with $N=4$
corresponds to photon-photon scattering and has a
potentially logarithmic divergence, but, as a consequence of
gauge invariance, it is actually convergent. Finally, using
Furry's theorem, we can discard the loop with $N=3$ (for
these topics see for instance refs. \cite{bogoliubov}).}. To
regularize the vacuum polarization diagram in a way which
does not break gauge invariance, we shall use the PV
regularization \cite{pauli}, for which some details will be
given in the next section. It is regularized introducing a
massive PV spinor field of bosonic type into the Lagrangian.

Indicating with $\psi$ and $\psi_1$ the physics and PV field
respectively, the Euclidean generating functional regularized to
the $M_0$ scale is {\sacolsep
\beaa
Z[J,A;M_0] & = & \ipsi \exp \bigg\{ - \frac{1}{4e^2_0}
\int_x F^2_{\mu\nu} + \int_x \pb (i\dslash - m_0)
\psi + \int_x \pb_1 (i\dslash - M_0) \psi_1 \nonumber \\[3mm]
&& + \int_x (\cb \psi + \pb \chi) +
\int_x (\cb_1 \psi_1 + \pb_1 \chi_1)\bigg\} \; .
\eeaa}
As $A_{\mu}$ is a classical field, we have not considered
gauge fixing terms. The dependence on $x$ has been
understood and
\be
\begin{array}{c}
\Psi \doteq \big\{\Phi; \Phi^{\textrm{\tiny{PV}}}\big\}
\doteq \big\{\psi, \pb; \psi_1, \pb_1\big\} \; , \\[4mm] J
\doteq \big\{\mathcal{J}; \mathcal{J}^{\textrm{\tiny{PV}}}\big\} \doteq
\big\{\cb, \chi; \cb_1, \chi_1\big\} \; ,
\end{array} \qquad \qquad
\begin{array}{c}
\mathcal{D} \Psi \doteq \mathcal{D} \psi \mathcal{D} \pb
\mathcal{D} \psi_1 \mathcal{D} \pb_1 \; , \\[4mm]
{\ds \int_x \doteq \int d^4x} \; .
\end{array}
\ee

Now Wilson's idea of RG is applied. It consists in lowering
the scale $M_0$ to $M$ while determining the effective
action which compensates for the loss of modes. As we shall
show, the functional {\sacolsep
\beaa \label{defstot}
Z[J,A;M,M_0] & = & \ipsi \exp \bigg\{ -M\int_x
\pb_1 \psi_1 - S_{\mathrm{eff}}[A,\Psi,\mathcal{J};M,M_0]
+ \int_x f_M(\cb_1 \psi_1 + \pb_1 \chi_1)\bigg\} \nonumber
\\[3mm] & \equiv & \ipsi \exp( - S_{\mathrm{tot}})
\eeaa}
is equal to $Z[J,A;M_0]$, except for a tree level two-point
function, provided that the effective action and $f_M$
satisfy proper RG equations with the initial conditions
{\sacolsep
\beaa \label{ciaw}
\lim_{M \to M_0} S_{\mathrm{eff}}[A,\Psi,\mathcal{J};M,M_0] &=&  \frac{1}{4e^2_0}
\int_x F^2_{\mu\nu} - \int_x \pb (i\dslash - m_0)\psi
\nonumber \\[3mm]&& - \int_x \pb_1 i\dslash \psi_1 - \int_x (\cb \psi + \pb
\chi) \; ,
\eeaa}
\be \label{cifm}
\lim_{M \to M_0} f_M = 1 \; .
\ee
From the RG point of view we have only renormalized the
sources associated with the fields which flow. Even if $f_M$
could be $x$-dependent in principle, it will be shown that
in fact it only depends on the scale.

The flow equations are obtained from a Polchinski identity:
{\sacolsep
\beaa \label{eqpol}
0 & = & \int_x M \frac{\partial}{\partial M}
\left(\frac{1}{M}\right) \ipsi \bigg\{\frac{\delta}{\delta
\psi_1}\left( \psi_1 M + \frac{1}{2}\frac{\delta}{\delta
\pb_1} \right) \nonumber \\[3mm] && + \frac{\delta}{\delta
\pb_1}\left( \pb_1 M + \frac{1}{2}\frac{\delta}{\delta
\psi_1} \right)\bigg\} \exp( - S_{\mathrm{tot}}) \; .
\eeaa}
This identity has been written directly in the $x$-space because
the role of the cut-off function is played by $1/M$, which does not
depend on the momenta. This is not true for conventional cut-off
methods. Indicating with
\be
\big<\mathcal{O}\big>=\ipsi \mathcal{O} \: \exp(-S_{\mathrm{tot}})
\ee
the quantum average of an operator $\mathcal{O}$ in an external
field $A_{\mu}$ and in presence of sources $J$, eq. (\ref{eqpol})
becomes
\be \label{eqstot}
0 = \int_x \bigg<\frac{\delta \psi_1}{\delta
\psi_1} + \frac{\delta \pb_1}{\delta \pb_1} - \frac{\delta S_{\mathrm{tot}}}
{\delta \psi_1} \psi_1 - \pb_1 \frac{\delta
S_{\mathrm{tot}}}{\delta
\pb_1} + \frac{1}{M} \bigg(\frac{\delta S_{\mathrm{tot}}}{\delta \psi_1}
\frac{\delta S_{\mathrm{tot}}}{\delta \pb_1} - \frac{\delta^2 S_{\mathrm{tot}}}{\delta
\psi_1 \delta \pb_1}\bigg)\bigg> \; .
\ee

We have come across the quantity
\be \label{vacuum}
\bigg<\frac{\delta \psi_1}{\delta \psi_1} +
\frac{\delta \pb_1}{\delta \pb_1}\bigg> \; .
\ee

Similar terms were discarded in ref. \cite{polchinski} and
in the following literature on RG with the exception of
refs. \cite{morris0}--\cite{yoshida}, \cite{yoshidac}. They
have been interpreted as ``Wilson lines biting their own
tails'' in the gauge invariant formulation of the exact RG
proposed by Morris \cite{morris0, morris12}. Showing their
anomalous origin when an external gauge field is
present\footnote{In \cite{yoshida, yoshidac} the external
gauge field is a component of a vector superfield.} and the
flow of regulator fields with respect to a mass parameter is
considered, the physical meaning of terms analogous to
(\ref{vacuum}) has been elucidated in refs. \cite{yoshida,
yoshidac}. From this point of view it is clear why
Polchinski could discard these terms: this is a legitimate
assumption for a theory like $\lambda \phi^4$ that does not
have a background gauge field\footnote{These terms
contribute to the partition function with a non-influential
overall factor.}. Following \cite{yoshida, yoshidac}, we
have to evaluate the quantity (\ref{vacuum}) carefully, as
it assumes the meaning of vacuum energy of $\psi_1$ and
$\pb_1$ when the external gauge field $A_{\mu}$ is present.
We shall calculate this quantity after the RG equations have
been obtained.

Substituting $S_{\mathrm{tot}}$ in eq. (\ref{eqstot}) for
the expression defined in (\ref{defstot}), we obtain
{\sacolsep
\beaa \label{eqprg}
0 & = & \bigg< \frac{1}{2} \int_x \left(\frac{\delta
\psi_1}{\delta \psi_1} + \frac{\delta \pb_1}{\delta \pb_1}\right) + \frac{1}{M}
\int_x \left(\frac{\delta S_{\mathrm{eff}}}{\delta \psi_1}
\frac{\delta S_{\mathrm{eff}}}{\delta \pb_1} - \frac{\delta^2 S_{\mathrm{eff}}}{\delta
\psi_1 \delta \pb_1}\right)  \nonumber \\[3mm] && - M \int_x \pb_1 \psi_1
+ \int_x f_M (\cb_1 \psi_1 + \pb_1 \chi_1) - \frac{1}{M}
\int_x f^2_M \cb_1 \chi_1\bigg> \; .
\eeaa}
Note that we have used
\be
\frac{\delta^2 S_{\mathrm{tot}}}{\delta \psi_1 \delta \pb_1} =
\frac{1}{2} \left(\frac{\delta^2 S_{\mathrm{tot}}}{\delta
\psi_1 \delta \pb_1} + \frac{\delta^2 S_{\mathrm{tot}}}{\delta
\pb_1 \delta \psi_1}\right) = \frac{M}{2} \left(\frac{\delta \psi_1}{\delta
\psi_1} + \frac{\delta \pb_1}{\delta \pb_1}\right) + \frac{\delta^2 S_{\mathrm{eff}}}
{\delta \psi_1 \delta \pb_1} \; ,
\ee
which tells us that the quantity ({\ref{vacuum}}) is also
due to the mass term of the PV field, and the equations
$\big<\delta S_{\mathrm{tot}}/\delta \psi_1\big> =
\big<\delta S_{\mathrm{tot}}/\delta\pb_1\big> = 0$. From a
comparison between eq. (\ref{eqprg}) and the $M$ derivative
of $Z[J,A;M,M_0]$, the physics is kept unchanged lowering
the scale if the following RG equations are satisfied:
\begin{enumerate}
\item RG equation for the effective action
\beaa \label{erge}
\lefteqn{\left<M \frac{\partial}{\partial M} \left\{S_{\mathrm{eff}} +
\frac{1}{2} \ln \left(\frac{M}{M_0}\right) \int_x
\left(\frac{\delta \psi_1}{\delta \psi_1} +
\frac{\delta \pb_1}{\delta \pb_1}\right)\right\}
\right>} \nonumber \\[3mm] && = - \frac{1}{M}
\int_x \left<\frac{\delta S_{\mathrm{eff}}}{\delta \psi_1}
\frac{\delta S_{\mathrm{eff}}}{\delta \pb_1} - \frac{\delta^2 S_{\mathrm{eff}}}{\delta
\psi_1 \delta \pb_1}\right>
\eeaa
with the initial condition (\ref{ciaw}).
\item RG equation for the support $f_M$
\be
M \frac{\partial f_M}{\partial M} = f_M
\ee
with the initial condition (\ref{cifm}), for which the solution is
obviously $f_M
= M/M_0$.
\end{enumerate}
In fact, it follows the condition of RG invariance:
\be
M \frac{\partial}{\partial M} \left\{\exp\left(-
\frac{M}{M^2_0} \int_x \cb_1 \chi_1\right)Z[J,A;M,M_0]\right\} = 0
\; .
\ee
Using the initial condition $\lim_{M \to M_0} Z[J,A;M,M_0] =
Z[J,A;M_0]$, which is a consequence of (\ref{ciaw}) and
(\ref{cifm}), the solution of the last equation is
\be
Z[J,A;M_0] = \exp\left\{\left(\frac{1}{M_0} -
\frac{M}{M^2_0}
\right)\int_x \cb_1 \chi_1\right\} Z[J,A;M,M_0] \; .
\ee

The quantity (\ref{vacuum}) is evaluated by using Fujikawa's
path integral approach to the anomalous Ward-Takahashi
identities. By the rescaling
\be \label{resctrans}
\begin{array}{l}
{\ds \psi_1 \longrightarrow \psi'_1 = e^{\alpha}
\psi_1 \; ,} \\[4mm] {\ds \pb_1 \longrightarrow \pb'_1 = e^{\alpha}
\pb_1 \; ,}
\end{array}
\ee
with $\alpha$ function of $x$, the measure of the functional
integral transforms as follows:
\be
\mathcal{D}\Psi \longrightarrow \mathcal{D} \Psi'
= \mathcal{D}\Psi \; \exp 2 \int_x \alpha \mathcal{A}_1 = \mathcal{D}\Psi \;
\exp \frac{1}{12\pi^2}\int_x \alpha F^2_{\mu\nu} \; .
\ee
We have used the result quoted in appendix \ref{appx2} and the
commutative nature of the PV field. The related anomalous
Ward-Takahashi identity is obtained by a variational derivative:
\be
0 = \frac{\delta}{\delta \alpha} Z[J,A;M,M_0]
\bigg\vert_{\alpha=0} = \left<
\frac{\delta S_{\mathrm{tot}}}{\delta \psi_1}\psi_1 + \pb_1
\frac{\delta S_{\mathrm{tot}}}{\delta \pb_1} - 2\mathcal{A}_1\right> \; .
\ee
On the other hand, the identity
\be
\ipsi \left\{ \frac{\delta}{\delta \psi_1} \left(\psi_1
e^{-S_{\mathrm{tot}}}
\right) + \frac{\delta}{\delta \pb_1} \left(\pb_1
e^{-S_{\mathrm{tot}}}
\right)\right\} = 0
\ee
turns out to be
\be
\left<\frac{\delta \psi_1}{\delta \psi_1} +
\frac{\delta \pb_1}{\delta \pb_1}\right> =
\left<\frac{\delta S_\mathrm{tot}}{\delta \psi_1} \psi_1 +
\pb_1 \frac{\delta S_\mathrm{tot}}{\delta \pb_1} \right> \nonumber \\[3mm]
 = \left<2\mathcal{A}_1\right> \; .
\ee
Finally, using the independence of $\mathcal{A}_1$ on the regulator
fields $\psi_1$ and $\pb_1$, eq. (\ref{erge}) becomes
\be \label{ergef}
M \frac{\partial \widetilde{S}_{\mathrm{eff}}}{\partial M} =
- \frac{1}{M} \int_x \bigg(\frac{\delta \widetilde{S}_{\mathrm{eff}}}{\delta \psi_1}
\frac{\delta \widetilde{S}_{\mathrm{eff}}}{\delta \pb_1} -
\frac{\delta^2 \widetilde{S}_{\mathrm{eff}}}
{\delta \psi_1 \delta \pb_1}\bigg) \; ,
\ee
where
\be \label{npart}
\widetilde{S}_{\mathrm{eff}} = S_{\mathrm{eff}} +
\frac{1}{24\pi^2} \ln \left(\frac{M}{M_0}\right)
\int_x F^2_{\mu\nu} \; .
\ee

Note that we have dealt with the anomaly equations in the
operator form and only after having evaluated the quantity
(\ref{vacuum}) have we left out the quantum expectation
value. In other words, following ref. \cite{yoshidac}, we
have passed from the weak to the strong form of Polchinski's
equation. This is an important point because only by working
with the Wilsonian effective action ($S_{\mathrm{eff}}$),
can the relevance of the rescaling anomaly for the low
energy theory be studied. In fact, while the 1PI effective
action is a $c$-number function of classical fields,
$S_{\mathrm{eff}}$ is an operator which retains quantum
fields that have not been integrated out yet and therefore
the correct Jacobian has to be taken into account after a
rescaling of the fields\footnote{It could be shown that the
anomalous term in eq. (\ref{npart}) is subtracted in the
formal transition from $\widetilde{S}_{\mathrm{eff}}$ to the
generator of connected Green's functions with an infrared
mass cut-off $M$ -- which is obtained with the integration
of eq. (\ref{ergef}) -- if the latter is correctly
normalized. It is a result of the classical nature of its
fields.}.

As in ref. \cite{yoshida} we have identified the normal
($\widetilde{S}_{\mathrm{eff}}$) and anomalous part of the
Wilsonian effective action. It is the latter that is
responsible for the rescaling of the electric charge. In
fact, the solution $\widetilde{S}_{\mathrm{eff}}
[A,\Psi\big\vert_{\Phi^{\textrm{\tiny{PV}}}=0},
\mathcal{J};M,M_0]$ of eq. (\ref{ergef}) -- in terms of which the low energy physics at
the momentum scale $p \sim M' \ll M < M_0$ is given --
varies rather slowly:
\be
\widetilde{S}_{\mathrm{eff}}[A,\Psi\big\vert_{\Phi^{\textrm{\tiny{PV}}}=0},
\mathcal{J};M,M_0] \simeq \widetilde{S}_{\mathrm{eff}}[A,
\Psi\big\vert_{\Phi^{\textrm{\tiny{PV}}}=0},\mathcal{J};M_0,M_0] + O(1/M,1/M_0) \; .
\ee
Using (\ref{npart}) and the initial condition (\ref{ciaw}), we
obtain {\sacolsep
\beaa \label{eqsemmo}
S_{\mathrm{eff}}[A,\Psi\big\vert_{\Phi^{\textrm{\tiny{PV}}}=0},
\mathcal{J};M,M_0] & \simeq & \frac{1}{4}
\left(\frac{1}{e^2_0} -
\frac{1}{6\pi^2}\ln \frac{M}{M_0}\right) \int_x F^2_{\mu\nu}
\nonumber \\[3mm] && - \int_x \pb (i\dslash - m_0)\psi - \int_x (\cb \psi + \pb \chi)
\; .
\eeaa}
If we set $\Phi = 0$, the term on the left-hand side of
(\ref{eqsemmo}) which has a $F^2_{\mu\nu}$ structure is
selected, giving
\be
\frac{1}{e^2(M)} = \frac{1}{e^2_0} -\frac{1}{6\pi^2} \ln \frac{M}{M_0}
\; ,
\ee
from which the well known result of the one-loop $\beta$-function
can be obtained.

\section{Regularized gauge invariant effective action}
\label{secgiea}

In this section we use the Slavnov regularization of gauge
theories \cite{slavnov1}, in the framework of the background
field method, to regularize the theory at the one-loop
level.

It is a well known fact that, as a consequence of the gauge
fixing process, we have to deal with non-gauge invariant
quantities in the intermediate stage of the calculation of
the $S$-matrix. The gauge invariance of physical quantities
is guaranteed if the renormalization procedure satisfies
Slavnov-Taylor identities. However, there is a method that
retains a residual gauge invariance so that background
Slavnov-Taylor identities are fulfilled automatically. This
is the background field method (see for instance
\cite{abbott}--\cite{bfm}).

First of all, closely following Abbott's paper
\cite{abbott}, we shall give a brief presentation of the
background field formalism that incorporates the matter.
Each field of the theory is considered as a sum of a
classical background part $\Phi^{\tinyb}_i =
\big\{A^a_{\mu}, c^a_{\tinyb},
\bar{c}^a_{\tinyb}, \psi^f_{\tinyb},
\pb^f_{\tinyb} \big\}$ and a quantum piece $\Phi_i=
\big\{Q^a_{\mu}, c^a, \bar{c}^a, \psi^f, \pb^f \big\}$,
which represents the quantum fluctuation around the
background field. Then, using the covariant $\alpha$
background gauge, the Euclidean generating functional of the
non-Abelian theory can be written as{\sacolsep
\beaa \label{egfna}
\widetilde{Z}[\mathcal{J},\Phi_{\tinyb}] & = & \iphi
\exp \bigg\{ - S_{\mathrm{YM}}(A+Q) - \frac{1}{2\alpha g^2_0} \int_x
(D_{\mu}Q_{\mu})^a (D_{\nu} Q_{\nu})^a \nonumber \\[3mm] &&
+ \int_x (\bar{c}_{\tinyb} + \bar{c})^a
D^{ab}_{\mu}D^{bd}_{\mu}(A+Q) (c_{\tinyb} + c)^d \nonumber
\\[3mm] &&  + \int_x (\pb_{\tinyb} + \pb)^f
\left[i\dslash(A+Q) - m^f_0\right](\psi_{\tinyb} + \psi)^f + \int_x
\mathcal{J}_i \Phi_i\bigg\} \; ,
\eeaa}
where
\be
\begin{array}{l}
{\ds \mathcal{L}_{\mathrm{YM}}(Q) = \frac{1}{4g^2_0}
F^a_{\mu\nu}(Q) F^a_{\mu\nu}(Q) \qquad \mathrm{with} \;\;
F^a_{\mu\nu}(Q) =
\partial_{\mu} Q^a_{\nu} - \partial_{\nu} Q^a_{\mu} +
f^{abc} Q^b_{\mu}Q^c_{\nu} \; ,} \\[6mm] {\ds D_{\mu} = D_{\mu}(A)
= \partial_{\mu} - iA_{\mu} = \partial_{\mu} - iA^a_{\mu} T^a \; ,}
\\[4mm] {\ds (D_{\mu}Q_{\nu})^a = D^{ab}_{\mu}Q^b_{\nu} = \partial_{\mu} Q^a_{\nu} +
f^{abc} A^b_{\mu}Q^c_{\nu} \; ,} \\[4mm] {\ds \mathcal{J}_i
= \Big\{j^a_{\mu},\etab^a,-\eta^a, \cb^f,-\chi^f\Big\}} \; .
\end{array}
\ee
Furthermore, the color indices have been suppressed and
$f=1,\ldots,N_f$ is a flavor index, where $N_f$ is the number of
quark flavors. The gauge group is of color $SU(N)$ with the
Hermitian generators that satisfy the typical relations $[T^a, T^b]
= i f^{abc} T^c$ of the Lie algebra and are normalized as follows:
\be
\Tr (T^a T^b) = t_2(R) \delta^{ab}\; .
\ee
$t_2(R)$ is the Dynkin index of the $R$ representation, for which
$t_2(A) = N$ and $t_2(N) = 1/2$ when the adjoint ($A$) and
fundamental ($N$) representation are considered.

The generating functional in (\ref{egfna}) has the
remarkable property of being invariant under simultaneous
infinitesimal transformations of the background
fields{\sacolsep
\beaa \label{gtra}
\delta A^a_{\mu} & = & (D_{\mu} \omega)^a \; , \nonumber \\[2mm]
\delta \psi^f_{\tinyb} & = & i \omega^a T^a \psi^f_{\tinyb} \; , \qquad \quad
\delta \pb^f_{\tinyb}  =  - i \pb^f_{\tinyb} \omega^a T^a \; , \\[2.5mm]
\delta c^a_{\tinyb} & = & f^{abc} c^b_{\tinyb} \omega^c \; ,
\qquad \quad \; \delta \bar{c}^a_{\tinyb} = f^{abc}
\bar{c}^b_{\tinyb} \omega^c \; , \nonumber
\eeaa}
and the sources{\sacolsep
\beaa \label{gtrs}
\delta j^a_{\mu} & = & f^{abc} j^b_{\mu}
\omega^c \; , \nonumber \\[2mm] \delta \cb^f & = & - i \cb^f \omega^a T^a \; ,
\qquad \quad \delta \chi^f =  i \omega^a T^a \chi^f  \; , \\[2mm] \delta
\etab^a_{\tinyb} & = & f^{abc} \etab^b_{\tinyb} \omega^c \; , \qquad \quad
\; \; \; \delta \eta^a_{\tinyb} = f^{abc} \eta^b_{\tinyb} \omega^c \; . \nonumber
\eeaa}
Moreover, its connected part
$\widetilde{W}[\mathcal{J},\Phi_{\tinyb}] = \ln
\widetilde{Z}[\mathcal{J},\Phi_{\tinyb}]$ is equal to the background gauge
invariant effective action $\widetilde{\Gamma}[0,\Phi_{\tinyb}]$ --
for which an equivalence proof of the background field quantization
method with the conventional one can be inferred from refs. in
\cite{ags} -- if the sources $\mathcal{J}$ are
$\Phi_{\tinyb}$-dependent in such a way that a generalized 't Hooft
equation
\be \label{eqthooft}
\frac{\delta \widetilde{W}}{\delta \Phi^{\tinyb}_i(x)} +
\int_y \frac{\delta \mathcal{J}_j(y)}{\delta \Phi^{\tinyb}_i(x)}
\frac{\delta \widetilde{W}}{\delta \mathcal{J}_j(y)} =
- (-1)^{\delta_i} \mathcal{J}_i(x)
\ee
is satisfied. This is demonstrated in appendix \ref{appx1}
generalizing the equivalence proof of 't Hooft's procedure
\cite{hooft} with that of Abbott's \cite{abbott}, when
fermions are incorporated into the theory. We have
introduced the fermionic number $\delta_i$ such that
$(-1)^{\delta_i}=1$ and $(-1)^{\delta_i}=-1$ for bosonic and
fermionic variables respectively\footnote{It accounts for
the commutation property of the variables involved. For
example $\Phi_i\Phi_j = (-1)^{\delta_i
\delta_j}\Phi_j\Phi_i$.}.

The background gauge invariance sets constraints on the
infinities that appear in
$\widetilde{\Gamma}[0,\Phi_{\tinyb}]$ (see refs.
\cite{abbott, weinberg}). They must take the following gauge
invariant form\footnote{For the time being, if we do not
indicate the dependence on gauge fields, it will mean that
we are considering the background gauge field dependence.
For example $F^a_{\mu\nu}
\doteq F^a_{\mu\nu}(A)$.}
\be
\widetilde{\Gamma}^{\infty}_0 = \int_x \Big\{C_1 (F^a_{\mu\nu})^2 +
C_2 \pb^f_{\tinyb} \dslash \psi^f_{\tinyb} + C_3
\pb^f_{\tinyb} \psi^f_{\tinyb} + C_4 (D_{\mu} \bar{c}_{\tinyb})^a
(D_{\mu} c_{\tinyb})^a \Big\} \; ,
\ee
where $C_n$, with $n=1,\ldots,4$, are infinite constants and
the lower index on the left-hand side means that we are
taking the bare quantities on the other side. In terms of
renormalized fields $\Phi^{\tinyb}_i = Z^{-1/2}_i
(\Phi^{\tinyb}_0)_i$ the last identity becomes
\be
\widetilde{\Gamma}^{\infty} = \int_x \Big\{C_1 Z^{1/2}_A(F^a_{\mu\nu})^2 +
C_2 Z_{\psi_{\tinyb}} \pb^f_{\tinyb} \dslash \psi^f_{\tinyb}
+ C_3 Z_{\psi_{\tinyb}} \pb^f_{\tinyb} \psi^f_{\tinyb} + C_4
Z_{c_{\tinyb}} (D_{\mu} \bar{c}_{\tinyb})^a (D_{\mu}
c_{\tinyb})^a \Big\} \; ,
\ee
with $F_{\mu\nu}$ and $D_{\mu}$ that will have the
expressions dictated by the gauge invariance if the constant
structure and the elements of the Lie algebra are
renormalized as follows: $f^{abc} = Z^{1/2}_A f^{abc}_0$ and
$T^a = Z^{1/2}_A T^a_0$. These quantities are determined by
the Lie algebra relations except for a multiplicative common
factor, which is the gauge coupling constant, if the Lie
algebra is simple. Thus, the gauge coupling must renormalize
as $g = Z^{1/2}_A g_0$, which means that the gauge coupling
and background gauge field renormalization are related. In
fact, defining $Z_g = g_0/g$, the relation $Z_g =
Z^{-1/2}_A$ is obtained, that is to say the $\beta$-function
originates solely from the background gauge field two-point
function \cite{abbott}. This is the reason why from now on
we shall be interested in the gauge invariant effective
action $\widetilde{\Gamma}[0,A] =
\widetilde{W}[\mathcal{J}[A],A]$, whose path integral
representation is deduced from eq. (\ref{egfna}) setting
$\psi^f_{\tinyb} = \pb^f_{\tinyb}
= c^a_{\tinyb} = \bar{c}^a_{\tinyb} = 0$, with the sources
$\mathcal{J}_i[A]$ which are solutions of suitable 't
Hooft's equations\footnote{They are obtained from eqs.
(\ref{eqthooft}) and (\ref{eqthooft2}) noting that the
condition $\widetilde{\Phi}_i=0$ is now equivalent to
$\ovp_i = \delta_{i1} A^a_{\mu}$ (see the procedure in
appendix \ref{appx1}).}.

With the intention to calculate the one-loop
$\beta$-function in the next section taking full advantage
of the gauge invariance, we regularize the functional
$\widetilde{\Gamma}[0,A]$ using the regularization proposed
by Slavnov \cite{slavnov1} at the one-loop level. It
consists of the following two steps. The first one is a
gauge invariant generalization of the higher derivatives
regularization. In fact, to improve the ultraviolet behavior
of propagators, the gauge invariance requires the
introduction of covariant instead of ordinary derivatives
into the kinetic term of the action \cite{slavnov0}. Thus,
the Yang-Mills action and the gauge fixing surface $(G^a)$
are replaced by the substitutions
\be \label{hcdr}
S_{\mathrm{YM}}(A+ Q)\longrightarrow
S^{n,\l}_{\mathrm{YM}}(A+Q)
= \frac{1}{4g^2_0} \int_x \left\{F^2_{\mu\nu} + \frac{1}{\l^{2n}} (D^n F_{\mu\nu})^2
\right\}(A+ Q) \; , \\[2mm]
\ee
\be \label{sfgagfix}
G^a = (D_{\mu} Q_{\mu})^a \longrightarrow F_n(D^2/\l^2) (D_{\mu}
Q_{\mu})^a \; ,
\ee
where $F_n$ is a polynomial of an order greater than equal
to $n/2$ and from now on $V^2 \doteq V^aV^a$.

For the reason mentioned above,
$S^{n,\l}_{\mathrm{YM}}(A+Q)$ is invariant under the quantum
gauge transformation $\delta (A+Q)^a_{\mu} =
\delta Q^a_{\mu} = D^{ab}_{\mu}(A+Q)\omega^b$.
Moreover, the substitutions (\ref{hcdr}) and
(\ref{sfgagfix}) yield a functional
$\widetilde{\Gamma}^n_{\l}[0,A]$ still invariant under the
background gauge transformation $\delta A^a_{\mu}
= (D_{\mu}\omega)^a$. Therefore, the advantages of
background field method are retained in the regularized
theory. For instance, the identity $Z_g = Z^{-1/2}_A$
remains true in the regularized theory. This is a
significant property, which is a result of using the
background field approach to the Slavnov regularization.

An inspection of the superficial degree of divergence of Feynman's
diagrams, with the classical field $A$ on external lines and
quantum fields $Q$, $c$, $\bar{c}$, $\psi$ and $\pb$ inside loops,
tells us that the infinities only appear at the one-loop level if
$n\geq 2$ \cite{slavnov0, slavnov1, falceto, slavnov2} and matter
loops are regularized by the conventional PV regularization
\cite{pauli}. The second step concerns the regularization of
remaining divergences using the gauge invariant PV procedure
extended to Yang-Mills and ghost loops \cite{slavnov1}.

The one-loop contribution to $\widetilde{\Gamma}^n_{\l}[0,A]$ is
given by the partition function{\sacolsep
\beaa \label{1loopf}
\mathcal{Z}^n_{\l}[A] & = & \exp \widetilde{\Gamma}^{1\textrm{-}\mathrm{loop}}_{n,\l}[0,A]
= \int \mathcal{D}Q \; \exp \bigg\{- \frac{1}{2} \int_{xy}
\frac{\delta^2 S^{n,\l}_{\mathrm{YM}}}{\delta A^a_{\mu}(x) \delta A^b_{\nu}(y)}
Q^a_{\mu}(x) Q^b_{\nu}(y) \nonumber \\[3mm] && -
\frac{1}{2\alpha g^2_0} \int_x \left[F_n(D^2/\l^2)
D_{\mu}Q_{\mu}\right]^2 \bigg\} \det(i\dslash - m^f_0)
\det\left[F_n(D^2/\l^2)D^2\right] \; ,
\eeaa}
whose divergences can be cured compatibly with background gauge
invariance adding mass terms to each quantum field. In fact, the
functional{\sacolsep
\beaa \label{detpv}
\mathcal{Z}^n_{\l,M_i,\mu_j,m_k}[A] & = & \mathcal{Z}^n_{\l}[A]
\prod^{N_f}_{f=1} \prod^{n_1,n_2,n_3}_{i,j,k=1} \mathrm{det}^{-\alpha_i/2}
\mathcal{Q} (A,M_i,F_n) \nonumber \\[3mm] && \times \mathrm{det}^{\beta_j}
\left[F_n(D^2/\l^2)D^2 - \mu^2_j\right] \,
\mathrm{det}^{\gamma_k} (i\dslash - m^f_k) \; ,
\eeaa}
where{\sacolsep
\beaa
\mathrm{det}^{-1/2} \mathcal{Q} (A,M,F_n)
& = & \int \mathcal{D}Q \; \exp \bigg\{- \frac{1}{2} \int_{xy}
\frac{\delta^2 S^{n,\l}_{\mathrm{YM}}}{\delta A^a_{\mu}(x) \delta A^b_{\nu}(y)}
Q^a_{\mu}(x) Q^b_{\nu}(y) \nonumber \\[3mm] && - \frac{1}{2\alpha
g^2_0} \int_x \left[F_n(D^2/\l^2) D_{\mu}Q_{\mu}\right]^2
- \frac{M^2}{2} \int_x Q^2_{\mu}\bigg\} \;,
\eeaa}
is gauge invariant and even free of divergences if the PV
conditions
\be \label{pvcond}
\begin{array}{c}
{\ds \sum^{n_1}_{i=0} \alpha_i = 0 \; , }\\[6mm] {\ds
\sum^{n_1}_{i=0} \alpha_i M^2_i = 0 \; ,}
\end{array} \qquad \qquad
\begin{array}{c}
{\ds \sum^{n_2}_{j=0} \beta_j = 0 \; ,} \\[6mm] {\ds
\sum^{n_2}_{j=0} \beta_j \mu^2_j = 0 \; ,}
\end{array} \qquad \qquad
\begin{array}{c}
{\ds \sum^{n_3}_{k=0} \gamma_k = 0 \; ,} \\[6mm] {\ds
\sum^{n_3}_{k=0} \gamma_k \big(m^f_k\big)^2 = 0 }
\end{array}
\ee
are satisfied. In these equations $\alpha_0 = \beta_0 =
\gamma_0 = 1$ and $M_0 = \mu_0 = 0$. Note that, at the
one-loop level, there is no need to introduce a
pre-regulator and change the PV conditions in order to solve
the overlapping divergences problem \cite{falceto}, which is
due to subdiagrams that are not regularized by the PV
procedure.

The coefficients $\alpha_i$ ($\beta_j$ and $\gamma_k$) must
be integers in order that they can be interpreted as the
number of PV vector (scalar and spinor) fields of the
regularized local Lagrangian, whose masses are $M_i$
($\mu_j$ and $m^f_k$). In this case the PV procedure amounts
to subtract from each kind of loop a sequence of analogous
loops, along which massive fields propagate, which transform
under the same representation as the homogeneous Lorentz
group of the physical field in the former loop. The PV
fields corresponding to $\alpha_i < 0$ and $\beta_j > 0$ are
of fermionic type and those corresponding to $\gamma_k < 0$
are bosonics. Therefore, they do not satisfy the
spin-statistic relation. However, the spin-statistic theorem
is not violated because, decoupling from the physical fields
when the mass regulators go to infinity, no PV regulator
field appears in the asymptotic states. It should be
mentioned that, in the regularization scheme we are using,
this is true in any $\alpha
\neq 0$ gauge. When $\alpha = 0$ the regulator fields do not
decouple completely. In fact, the Landau gauge does not give
the correct value of the one-loop $\beta$-function of the
pure Yang-Mills theory as has been shown in ref.
\cite{martin}, which lead the authors to state a no-go
theorem concerning the Slavnov regularization. Even if a
minor modification of the scheme exists \cite{falceto}, in
which the correctness of the one-loop result on the
$\beta$-function is guaranteed, for the rest of this paper
we shall assume an $\alpha \neq 0$ gauge.

The Slavnov regularization does not specify the PV regulator
system. The only reasonable requirements that have to be
satisfied in addition to the conditions (\ref{pvcond}) are
the following. The coefficients $\alpha_i$, $\beta_j$ and
$\gamma_k$ must be chosen as integers. The variability field
of the mass regulators $M_i$, $\mu_j$ and $m^f_k$ has to
include infinity, which corresponds to the removal of the PV
part of the regularization. One of the different systems of
PV regulator fields is sufficient to calculate the one-loop
$\beta$-function with the RG method. However, it is worth
checking the independence of the one-loop $\beta$-function
on the PV regulator system.
\begin{figure}[t]
\hspace{0.8cm}
\includegraphics[width=17cm]{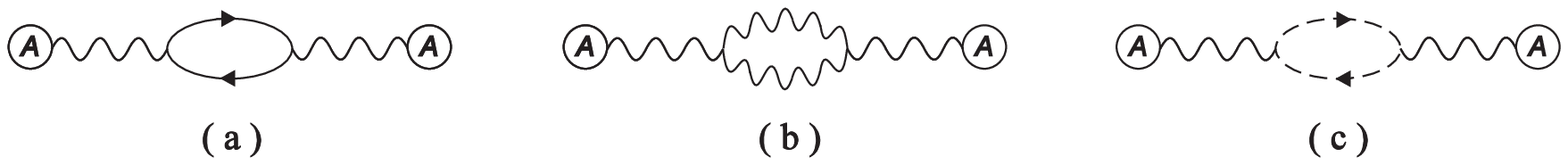}
\caption{\label{fig} \small{One-loop diagrams with the background gauge field $A$ on
external lines. Internal wavy lines are quantum gauge
propagators and dashed lines are ghost propagators.}}
\end{figure}

A suitable system for the calculation of the one-loop
$\beta$-function can be deduced from the relation $Z_g
= Z^{-1/2}_A$. From the knowledge of $Z_A$, for which
only the background field two-point function is required, we
can determine the $\beta$-function. Therefore, no vertex
function or tadpole diagram need to be considered, and, at
the one-loop level, we only need to regularize the Feynman
diagrams in figure \ref{fig}. The vacuum polarization
diagram (a) can be regularized as in section \ref{secba}.
Then, there is only one class of bosonic fields $\psi^f_1$
with mass $M$ in the spinor sector of the PV regulator
system. The diagrams (b) and (c) in figure \ref{fig} are
regularized if the usual PV conditions are satisfied. These
conditions can be realized through the introduction of at
least two auxiliary masses. In such a case we find
\be
\alpha_1 = \frac{M^2_2}{M^2_1 - M^2_2} \; , \qquad \quad
\alpha_2 = \frac{M^2_1}{M^2_2 - M^2_1}
\ee
for PV vector fields and the same for PV scalar fields
replacing $\alpha_i\big\vert_{i=1,2}$ and
$M_i\big\vert_{i=1,2}$ with $\beta_j\big\vert_{j=1,2}$ and
$\mu_j\big\vert_{j=1,2}$ respectively. Choosing the integer
values $\alpha_1=\beta_1=1$, it follows
$\alpha_2=\beta_2=-2$. Thus, the PV vectorial sector is
composed of one class of bosonic fields $Q^a_{1,\mu}$ of
mass $M_1$ and two mass degenerate classes of fermionic
fields $Q^a_{2,\mu}$ and $Q^a_{3,\mu}$ with mass
$M_2=M_1/\sqrt{2}$; the scalar sector of a set of fields
$c^a_{1,\mu}$, $c^a_{2,\mu}$ and $c^a_{3,\mu}$ with opposite
statistics and masses $\mu_1$ and
$\mu_2=\mu_3=\mu_1/\sqrt{2}$. Noting that the vector, scalar
and spinor loops are regularized separately, we can set
$M_2=\mu_2=M$.

Another system of PV regulator fields is inferred from the
chiral gauge invariant PV regularization proposed by Frolov
and Slavnov \cite{frolov}. Each ``sector'' is composed of an
infinite number of fields with alternating statistics, which
corresponds to the choice of $\alpha_i=\beta_i=\gamma_i=
(-1)^i$ for $i=\pm 1,\pm 2, \ldots,
\pm \infty$. However, a fundamental issue is to define how to sum over the
infinite number in order to satisfy the first PV condition.
In other words, we have to define the symbol
$\sum^{+\infty}_{n
= -\infty} (-1)^n$, which is a divergent series\footnote{We follow
Hardy's criterion \cite{hardy} to term a divergent series
the one that does not converge according to the classical
definition of Cauchy.}. There are various methods of summing
divergent series as part of the theory of divergent series,
for which we shall refer to Hardy's book \cite{hardy}. Then,
the series $\sum^{\infty}_{n=0} (-1)^n$ being Ces\`aro, Abel
and Euler summable to $1/2$, the following
manipulations{\sacolsep
\beaa \label{hardys}
0 & = & \sum^{+\infty}_{n = 0} (-1)^n - \sum^{+\infty}_{n = 0}
(-1)^n = \sum^{+\infty}_{n = 0} (-1)^n + \sum^{+\infty}_{n = 0}
(-1)^{n + 1} = \sum^{+\infty}_{n = 0} (-1)^n + \sum^{+\infty}_{n =
1} (-1)^n \nonumber \\[3mm] & = & \sum^{+\infty}_{n = 0} (-1)^n +
\sum^{-1}_{n = - \infty} (-1)^n = \sum^{+\infty}_{n = - \infty} (-1)^n
\eeaa}
are correct. Therefore, the first PV condition is satisfied
with respect to the criteria of summability mentioned above.

The physical meaning underlying this mathematical topic is
as follows. We attempt to subtract the divergence of the
physical sector, which is selected by $i=j=k=0$, introducing
a pair of PV fields with the same statistic. This is
equivalent to subtracting the divergence twice, since we are
considering $-1+1-1=-1$. To remove this divergence we need
to introduce another pair with the opposite statistic of the
former, which yields $+1-1+1-1+1=+1$. This argument makes
clear that it is hopeless trying to regularize the theory by
a finite number of PV fields with alternating statistics.
Then, an infinite number is introduced giving the
possibility to reiterate to infinity the above steps until
the divergence is removed \cite{nittoh}. However, in our
opinion, it is necessary to assign a specific meaning to the
symbol $\sum^{+\infty}_{n =
-\infty} (-1)^n$ by making use of the divergent series theory.

The second PV condition can be formally satisfied with a
proper system of mass regulators. In fact, using the Leibniz
series
\be
\begin{array}{lll}
{\ds \sum^{+\infty}_{n=1} \frac{1}{n^2} = \frac{\pi^2}{6}} \qquad &
\longrightarrow & \qquad {\ds \sum^{+\infty}_{n=-\infty}{}\!\!\! ' \,\,\, \frac{1}{n^2} =
\frac{\pi^2}{3} \; ,} \\[6mm] {\ds \sum^{+\infty}_{n=1} \frac{(-1)^{n-1}}
{n^2} = \frac{\pi^2}{12}} \qquad & \longrightarrow & \qquad {\ds
\sum^{+\infty}_{n=-\infty}{}\!\!\! ' \,\,\, \frac{(-1)^n}{n^2} = - \frac{\pi^2}{6} \; ,}
\end{array}
\ee
where $\sum{}'_n \doteq \sum_{n \neq 0}$, we
obtain{\sacolsep
\beaa
0 & = & \sum^{+\infty}_{n=-\infty}{}\!\!\! ' \,\,\,
\frac{(-1)^n}{n^2} + \frac{\pi^2}{6} =
\sum^{+\infty}_{n=-\infty}{}\!\!\! ' \,\,\,
\frac{(-1)^n}{n^2} + \frac{1}{2} \sum^{+\infty}_{n=-\infty}{}\!\!\! '
\,\,\, \frac{1}{n^2} \nonumber \\[3mm] & = & \sum^{+\infty}_{n=-\infty}{}\!\!\! '
\,\,\, \frac{(-1)^n}{n^2} \left[1 + \frac{(-1)^n}{2}\right] \; .
\eeaa}
Then, remembering that each kind of loop is regularized
separately, we can set $M^2_n = \mu^2_n = M^2 u_n$ and
$(m^f_n)^2 = M^2 v^f_n$ with
\be
u_n = \frac{1}{n^2} \left[1 + \frac{(-1)^n}{2}\right] \, ,
\qquad v^f_n = \frac{1}{n^2} \left[1 + \frac{(-1)^n}{2} +
\frac{6(m^f_0)^2}{\pi^2}\right] \, ,
\ee
to satisfy the second PV conditions. Note that $u_n, v^f_n
> 0 $ $\forall n \neq 0$. Therefore, choosing $M_n = \mu_n = M \sqrt{u_n}$
and $m^f_n = M (v^f_n)^{1/2}$, the removal of the PV
regularization is given by $M \to \infty$.

The proof that this generalized PV regularization works is given in
ref. \cite{nittoh1} for the pure Yang-Mills theory taking as mass
regulator $M_i = M \vert i \vert$ and $\mu_j=\mu
\vert j \vert$ $\forall i,j$. We think that it can be extended to
include the matter taking $m^f_k = m \vert k \vert$ for $k
\neq 0$ and even to the background field formalism using the
tools developed in \cite{pronin}. The higher covariant
derivatives complicate the Feynman rules and hence make the
above proofs a difficult task. However, as will become
clearer in the next section, these complications could be
avoided in the calculation of the one-loop $\beta$-function.
In fact, if we had used the relation $Z_g = Z^{-1/2}_A$ from
the beginning, we would have had to regularize solely the
graphs of figure \ref{fig} introducing only PV regulator
fields, which does not spoil the gauge invariance of
$\widetilde{\Gamma}[0,A]$. Then, using the Feynman rules
derived by Abbott in ref. \cite{abbott}, the leading
divergence of diagrams (b) and (c) would have been given
by\footnote{As usual, the diagram (a) could be regularized
using one bosonic PV field.}{\sacolsep
\beaa
\int_p p^2 \sum^{+\infty}_{n = - \infty} \frac{(-1)^n}{(p^2 + M^2
n^2)^2} & = & - \int_p p^2 \frac{\partial}{\partial p^2}
\sum^{+\infty}_{n = - \infty} \frac{(-1)^n}{p^2 + M^2 n^2} \nonumber
\\[3mm] & = & - \int_p p^2 \frac{\partial}{\partial p^2} \frac{\pi}{M p}
\frac{1}{\sinh (\pi p/M)} \; ,
\eeaa}
which prove their finiteness for finite $M$. The calculation
performed in the next section is a first step towards using
the Slavnov regularization in the RG context.

\section{One-loop $\beta$-function of QCD}
\label{secbna}

Due to the large number of fields involved, we need a more
concise notation than the one adopted in section
\ref{secba}. All fields and sources are collected in the
column vectors
\be \label{columnv}
\mathbf{\Psi} = \left( \begin{array}{c}
\mathbf{\Phi}\\ [2mm]
\mathbf{\Phi}^{\textrm{\tiny{PV}}}
\end{array} \right)  \; , \qquad
\mathbf{J}=\left( \begin{array}{c}
\mathbf{\mathcal{J}} \\ [2mm]
\mathbf{\mathcal{J}}^{\textrm{\tiny{PV}}} \end{array} \right) \; ,
\ee
with
\be
\begin{array}{l}
\left( \begin{array}{c} Q^a_{i,\mu}/\sqrt{2} \\ [2mm] Q^a_{i,\mu}/\sqrt{2} \\ [2mm]
c^a_j \\ [2mm] \bar{c}^a_j \\ [2mm] \psi^f_k \\ [2mm]
\big(\pb^f_k\big)^T \end{array} \right) =
\left\{ \begin{array}{ll} \mathbf{\Phi} & \qquad \qquad \mathrm{for} \;\;
i=j=k=0 \\[6mm] \mathbf{\Phi}^{\textrm{\tiny{PV}}} & \qquad \qquad
\mathrm{for} \;\; i,j,k \neq 0 \end{array} \right.
\end{array}
\ee
and, using the fermionic number introduced in the previous section,
\be
\begin{array}{l}
\left( \begin{array}{c} j^a_{i,\mu}/\sqrt{2} \\ [2mm] j^a_{i,\mu}/\sqrt{2} \\ [2mm]
\etab^a_j \\ [2mm] (-1)^{\delta_j} \eta^a_j \\ [2mm] \big(\cb^f\big)^T_k \\ [2mm]
(-1)^{\delta_k} \chi^f_k
\end{array}
\right) =
\left\{ \begin{array}{ll} \mathbf{\mathcal{J}} & \qquad \qquad \mathrm{for} \;\;
i=j=k=0 \\[6mm] \mathbf{\mathcal{J}}^{\textrm{\tiny{PV}}} & \qquad
\qquad
\mathrm{for} \;\; i,j,k \neq 0 \end{array} \right. \; .
\end{array}
\ee
We have considered $\pb$ a row vector following the Dirac
formalism and the vectorial sector has been doubled to treat
it as the scalar and spinor sector. This does not mean that
the respective measure in the integral functional that will
be considered below doubles, i.e.
\be
\mathcal{D}\Psi \equiv \prod_{i,j,k,f} \mathcal{D} Q_i
\mathcal{D} c_j \mathcal{D} \bar{c}_j
\mathcal{D} \psi^f_k \mathcal{D} \pb^f_k \; .
\ee
The generator functional of QCD, regularized according to
the Slavnov regularization in the framework of the
background field method, is the following:{\sacolsep
\beaa
\widetilde{Z} \big[\mathcal{J}(A),A,\jpv;M_0,\l_0\big] & =
& \exp \widetilde{\Gamma}\big[0,A, \jpv ;M_0,\l_0\big]
\nonumber \\[3mm] & = & \ipsi \exp \bigg\{- \frac{1}{2}
\big(\phipv, \M_0 \phipv \big) \nonumber \\[3mm] && - S_{\mathrm{int}}
[\mathcal{J}(A), A, \Psi; M_0,\l_0] + \big(\jpv, \phipv \big)\bigg\}
\;.
\eeaa}
The notation has been misused in calling{\sacolsep
\beaa
&& S^{n,\l_0}_{\mathrm{YM}}(A+Q) + \frac{1}{2} \int_{xy}
\frac{\delta^2 S^{n,\l_0}_{\mathrm{YM}}}{\delta A^a_{\mu}(x) \delta A^b_{\nu}(y)}
Q^a_{i,\mu}(x) Q^b_{i,\nu}(y) \Big \vert_{i \neq 0} +
\frac{1}{2\alpha g^2_0} \int_x \left[F_n(D^2/\l^2_0) D_{\mu}Q_{i,\mu}\right]^2
\nonumber \\[3mm] && - \int_x \bar{c}_j F_n(D^2/\l^2_0) D^2 c_j +
i\int_x \bar{c} F_n(D^2/\l^2_0) D_{\mu} Q_{\mu} c \nonumber
\\[3mm] && - \int_x \pb^f_k i \dslash \psi^f_k - \int_x
\pb^f \big(\Qslash - m^f_0\big) \psi^f - \big(\mathcal{J}(A), \Phi \big)
\eeaa}
the interaction action $S_{\mathrm{int}}[\mathcal{J}(A),
A,\Psi; M_0,\l_0]$, and
\be
(\Psi, \mathcal{A} \Psi) = \int_x \mathbf{\Psi}^T \mathbf{
\mathcal{A}} \mathbf{\Psi} = \int_x \Psi^T_{\alpha}
\mathcal{A}_{\alpha \beta} \Psi_{\beta}
\ee
denotes the inner product in the space spanned by the vector
$\mathbf{\Psi}$, where $\mathbf{\mathcal{A}}$ is a generic matrix.
Obviously, $\frac{1}{2} \big(\phipv, \M_0 \phipv
\big)$ is an inner product in the PV subspace. The mass
matrix is
\be
\mathbf{\M} = \left(\begin{array}{ccccccc}
0 & (-1)^{\delta_i} M^2_i & 0 & 0 & 0 & 0 \\[2mm] M^2_i & 0 & 0 & 0
& 0 & 0 \\[2mm] 0 & 0 & 0 & (-1)^{\delta_j} \mu^2_j & 0 & 0
\\[2mm] 0 & 0 & \mu^2_j & 0 & 0 & 0 \\[2mm] 0 & 0 & 0 & 0 &
0 & (-1)^{\delta_k}m^f_k \\[2mm] 0 & 0 & 0 & 0 & m^f_k & 0
\end{array}\right) \; ,
\ee
with $i,j,k \neq 0$. The matrix depends on the system of PV
regulator fields. The system described in the previous
section with a finite number of fields results in the
$(14\times14)$ matrix whose elements are given by setting
$M_2=M_3=\mu_2=\mu_3=m^f_1=M$, $M_1=\mu_1=\sqrt{2}M$,
$\delta_i\big\vert_{i=1} = \delta_j
\big\vert_{j=2,3} = 0$, $\delta_i\big\vert_{i=2,3} = \delta_j\big\vert_{j=1}=1$
and $\delta_k \big\vert_{k=1} = 0$. The one with an infinite
number results in the $(\infty\times\infty)$ matrix whose
elements are given by setting
$\delta_i\vert_{i=\mathrm{even}} = 0$,
$\delta_i\big\vert_{i=\mathrm{odd}} = 1$, $\delta_j,\delta_k
\big\vert_{j,k = \mathrm{even}} = 1$, $\delta_j,\delta_k \big\vert_{j,k =
\mathrm{odd}} = 0$ and  $M_n = \mu_n = M \sqrt{u_n}$, $m^f_n = M (v^f_n)^{1/2}$
or $M_i = \mu_i = m^f_i = M \vert i \vert$. The matrix $\M_0$ is
$\M$ to the scale $M_0$, namely $\M_0 \equiv \M (M \to M_0)$.

As in section \ref{secba} we vary the mass parameter $M_0$
to a lower value $M$ while keeping the physics unchanged. In
other words, we look for the RG equations, with the initial
conditions
\be \label{cisqcd}
\lim_{M \to M_0} S_{\mathrm{eff}}[\mathcal{J}(A), A, \Psi; M, M_0,\l_0] =
S_{\mathrm{int}}[\mathcal{J}(A), A, \Psi; M_0,\l_0] \; ,
\ee
\be \label{cijqcd}
\lim_{M \to M_0} \jtpv = \jpv \; ,
\ee
that have to satisfy $S_{\mathrm{eff}}$ and $\jtpv$ in order
that{\sacolsep
\beaa \label{mgfnab}
\widetilde{Z}\big[\mathcal{J}(A),A,\jtpv;M,M_0,\l_0\big] & = &
\ipsi \exp \bigg\{- \frac{1}{2} \big(\phipv, \M \phipv \big) \nonumber \\[3mm] &&
- S_{\mathrm{eff}}[\mathcal{J}(A), A, \Psi; M, M_0,\l_0] + \big(\jtpv, \phipv \big)\bigg\}
\nonumber \\[3mm] & \equiv & \ipsi \exp ( - S_{\mathrm{tot}})
\eeaa}
is equal to
$\widetilde{Z}\big[\mathcal{J}(A),A,\jpv;M_0,\l_0\big]$
except for a tree level two-point function. Closely
following the Abelian case, by means of the Polchinski
identity
\be \label{polinab}
0 = \int_x (-1)^{\delta_{\alpha}} M \frac{\partial
\M^{-1}_{\alpha\beta}}{\partial M} \ipsi \frac{\delta}{\delta \phipv_{\alpha}}
\bigg\{\M_{\beta\gamma} \phipv_{\gamma} + \frac{1}{2}\frac{\delta}{\delta
(\phipv)^T_{\beta}} \bigg\} \exp(- S_{\mathrm{tot}})
\ee
we obtain{\sacolsep
\beaa \label{eqpinab}
0 & = & \int_x \bigg<(-1)^{\delta_{\alpha}} M \frac{\partial
\M^{-1}_{\alpha\beta}}{\partial M} \M_{\beta\gamma}
\frac{\delta \phipv_{\gamma}}{\delta \phipv_{\alpha}} -
(-1)^{\delta_{\alpha}} M\frac{\partial
\M^{-1}_{\alpha\beta}}{\partial M} \M_{\beta\gamma}
\frac{\delta  S_{\mathrm{tot}}}{\delta \phipv_{\alpha}}
\phipv_{\gamma} \nonumber \\[3mm] && + \frac{(-1)^{\delta_{\alpha}}}{2}
M \frac{\partial \M^{-1}_{\alpha\beta}}{\partial M}
\bigg(\frac{\delta S_{\mathrm{tot}}} {\delta \phipv_{\alpha}}
\frac{\delta S_{\mathrm{tot}}}{\delta (\phipv)^T_{\beta}} -
\frac{\delta^2 S_{\mathrm{tot}}}{\delta
\phipv_{\alpha}
\delta (\phipv)^T_{\beta}}\bigg)\bigg> \; .
\eeaa}

To go further we need the following properties of the mass matrix
\be
\M_ {\alpha\beta} = (-1)^{\delta_{\alpha} \delta_{\beta}} \M_ {\beta\alpha} =
(-1)^{\delta_{\alpha}} \M_ {\beta\alpha} = (-1)^{\delta_{\beta}}
\M_ {\beta\alpha} \; .
\ee
In this paper the fermionic number is never summed over the
repeated indices. For example, $(-1)^{\delta_{\rho}}
\M_{\rho\alpha} \M^{-1}_{\rho\beta} \doteq (-1)^{\delta_1} \M_{1\alpha}
\M^{-1}_{1\beta} + (-1)^{\delta_2} \M_{2\alpha}
\M^{-1}_{2\beta} + \ldots = \delta_{\alpha \beta}$.

Then, substituting $S_{\mathrm{tot}}$ in eq. (\ref{eqpinab})
for the expression defined in (\ref{mgfnab}) and using the
equation $\big<\delta S_{\mathrm{tot}}/\delta
\phipv_{\alpha}\big> = 0$, we get {\sacolsep
\beaa
0 & = & \bigg< \frac{1}{2} \left(\phipv, M \frac{\partial
\M}{\partial M} \phipv \right)- \frac{1}{2} \int_x
(-1)^{\delta_{\alpha}} \M^{-1}_{\alpha\beta} M \frac{\partial
\M_{\beta\gamma}}{\partial M} \frac{\delta \phipv_{\gamma}}{\delta \phipv_{\alpha}}
\nonumber \\[3mm] && + \frac{(-1)^{\delta_{\alpha}}}{2} M\frac{\partial
\M^{-1}_{\alpha\beta}}{\partial M} \int_x \bigg(\frac{\delta S_{\mathrm{eff}}}
{\delta \phipv_{\alpha}} \frac{\delta
S_{\mathrm{eff}}}{\delta (\phipv)^T_{\beta}} -
\frac{\delta^2 S_{\mathrm{eff}}}{\delta
\phipv_{\alpha} \delta (\phipv)^T_{\beta}}\bigg) \nonumber \\[3mm]
&& - \left(M \frac{\partial \M} {\partial M} \M^{-1} \jtpv ,
\phipv\right) - \frac{1}{2} \left(\jtpv, (-1)^{\delta} M
\frac{\partial \M^{-1}}{\partial M} \jtpv \right)\bigg> \; .
\eeaa}
It is an easy task to check the $M$ independence of the matrix
$\mathbf{\M}^{-1} M \partial \mathbf{\M}/\partial M$ for all PV
systems considered before. In fact, it turns out to be
\be
\mathbf{\M}^{-1} M \frac{\partial \mathbf{\M}}{\partial M} =
\left(\begin{array}{ccccccc}
2 & 0 & 0 & 0 & 0 & 0 \\ 0 & 2 & 0 & 0 & 0 & 0 \\ 0 & 0 & 2
& 0 & 0 & 0 \\ 0 & 0 & 0 & 2 & 0 & 0
\\ 0 & 0 & 0 & 0 & 1 & 0 \\ 0 & 0 & 0 & 0 & 0 & 1
\end{array}\right) \equiv \mathbf{\Delta} \; .
\ee
If the following RG equations are satisfied:
\begin{enumerate}
\item RG equation for the effective action
\beaa \label{ergenab}
\lefteqn{\left<M \frac{\partial}{\partial M} \left\{S_{\mathrm{eff}} +
\frac{1}{2} \ln \left(\frac{M}{M_0}\right) \int_x
(-1)^{\delta_{\alpha}} \Delta_{\alpha\beta} \frac{\delta
\phipv_{\beta}}{\delta \phipv_{\alpha}} \right\} \right>} \nonumber \\[3mm]
&& = \frac{(-1)^{\delta_{\alpha}}}{2} M\frac{\partial
\M^{-1}_{\alpha\beta}}{\partial M} \int_x
\bigg<\frac{\delta S_{\mathrm{eff}}} {\delta \phipv_{\alpha}}
\frac{\delta S_{\mathrm{eff}}}{\delta (\phipv)^T_{\beta}} -
\frac{\delta^2 S_{\mathrm{eff}}}{\delta
\phipv_{\alpha} \delta (\phipv)^T_{\beta}}\bigg>
\eeaa
with the initial condition (\ref{cisqcd}).
\item RG equations for the sources
\be
M \frac{\partial \mathbf{\M}}{\partial M} \mathbf{\M}^{-1}
\mathbf{\tilde{\mathcal{J}}}^{\textrm{\tiny{PV}}} =
M \frac{\partial \mathbf{\tilde{\mathcal{J}}}^{\textrm{\tiny{PV}}}}
{\partial M}
\ee
with the initial condition (\ref{cijqcd}),
\end{enumerate}
the condition of RG invariance
\be \label{rginvest}
M \frac{\partial}{\partial M} \left\{\widetilde{Z}_M
\exp \frac{1}{2} \int^{M}_{M_0} d M' \left(\jtpv, (-1)^{\delta}
M \frac{\partial \M^{-1}}{\partial M} \jtpv \right) (M \to M')
\right\} = 0
\ee
is established. In eq. (\ref{rginvest}) and from now on
$\widetilde{Z}_M \doteq \widetilde{Z}\big[\mathcal{J}(A) ,A
,\jtpv ; M,M_0,\l_0 \big]$.

The vacuum energy of the PV regulator fields is deduced from the
anomalous Ward-Takahashi identity related to the infinitesimal
rescaling
\be
\delta \mathbf{\Phi}^{\textrm{\tiny{PV}}} =
\delta \alpha \mathbf{\Delta} \mathbf{\Phi}^{\textrm{\tiny{PV}}} \; .
\ee
In fact, the measure of the functional integral transforms into
$\mathcal{D} \Psi' = J \mathcal{D} \Psi$ and hence
\be
0 = \frac{\delta \widetilde{Z}_M}{\delta (\delta \alpha)}
\bigg\vert_{\delta \alpha=0} =
\left<\Delta_{\alpha\beta} \phipv_{\beta} \frac{\delta S_{\mathrm{tot}}}
{\delta \phipv_{\alpha}} - \frac{\delta \ln J}{\delta (\delta
\alpha)}\right> \; ,
\ee
from which, using the identity
\be
\ipsi \frac{\delta}{\delta \phipv_{\alpha}}\left\{(-1)^{\delta_{\alpha}}
\Delta_{\alpha\beta} \phipv_{\beta} e^{-S_{\mathrm{tot}}}\right\} \; ,
\ee
we obtain
\be
\left<(-1)^{\delta_{\alpha}} \Delta_{\alpha\beta} \frac{\delta
\phipv_{\beta}}{\delta \phipv_{\alpha}} \right> = \left<\Delta_{\alpha\beta}
\phipv_{\beta} \frac{\delta S_{\mathrm{tot}}}{\delta \phipv_{\alpha}} \right> =
\left<\frac{\delta \ln J}{\delta (\delta \alpha)}\right> \; .
\ee
As shown in appendix \ref{appx2} the Jacobian $J$ does not depend
on $\phipv$. Therefore, taking the strong form of Polchinski's
equation, one gets
\be
M \frac{\partial \widetilde{S}_{\mathrm{eff}}}{\partial M}
= \frac{(-1)^{\delta_{\alpha}}}{2} M\frac{\partial
\M^{-1}_{\alpha\beta}}{\partial M} \int_x
\bigg(\frac{\delta \widetilde{S}_{\mathrm{eff}}} {\delta \phipv_{\alpha}}
\frac{\delta \widetilde{S}_{\mathrm{eff}}}{\delta (\phipv)^T_{\beta}} -
\frac{\delta^2 \widetilde{S}_{\mathrm{eff}}}{\delta
\phipv_{\alpha} \delta (\phipv)^T_{\beta}}\bigg) \; ,
\ee
where
\be
\widetilde{S}_{\mathrm{eff}} = S_{\mathrm{eff}} + \frac{1}{2} \ln \left(\frac{M}{M_0}
\right) \int_x \frac{\delta \ln J}{\delta (\delta \alpha)} \; .
\ee

Now, we want to show the equality of $\ln J$ for all PV
regulator systems considered above and that it has the
correct value to achieve the one-loop $\beta$-function. We
start with the system having a finite number of fields. The
vector and scalar fields are in the adjoint representation
of the $SU(N)$ gauge group; the spinor fields in the
fundamental one. Then, using the results quoted in appendix
\ref{appx2} and the proper statistic of the fields involved,
we obtain{\sacolsep
\beaa \label{logj}
\ln J & = & \int_x \delta \alpha \left(2\mathcal{A}_3 - 2 \mathcal{A}_3 -2 \mathcal{A}_3
-4 \mathcal{A}_2 + 4 \mathcal{A}_2 + 4 \mathcal{A}_2 + 2 N_f \mathcal{A}_1\right)
\nonumber \\[2mm] & = & \int_x \delta \alpha \left\{- \frac{11}{48\pi^2} t_2(A)
+ \frac{N_f}{12\pi^2} t_2(N)\right\} F^2_{\mu\nu}\;.
\eeaa}

The analysis of the system with an infinite number of fields
requires a clarification. We have satisfied the first PV conditions
by making use of some summability criteria of the divergent series
theory. It corresponds to assign a fixed order to the infinite
products in the functional measure, as can be made clear looking at
eq. (\ref{detpv}) and thinking how the PV conditions (\ref{pvcond})
come out. Taking as example the spinor sector, if{\sacolsep
\beaa
\mathrm{det} (i\dslash - m^f_0) \prod_k && \mathrm{det}^{\gamma_k}
(i\dslash - m^f_k) \qquad \longrightarrow \qquad \prod^{+
\infty}_{k = - \infty} \mathrm{det}^{(-1)^k} (i\dslash - m^f_k)\nonumber \\[3mm] &&
= \int \prod^{+ \infty}_{k = - \infty}
\mathcal{D}\psi^f_k \mathcal{D}\pb^f_k \; \exp
\sum^{+ \infty}_{k = - \infty} \int_x \pb^f_k  (i\dslash - m^f_k) \psi^f_k
\;,
\eeaa}
the first PV condition becomes $\sum^{+ \infty}_{k = - \infty}
(-1)^k = 0$. Thus, the measure functional is
\be
\mathcal{D}\Psi = \prod^{+ \infty}_{i = - \infty} \mathcal{D} Q_i
\prod^{+ \infty}_{j = - \infty} \mathcal{D} c_j \mathcal{D} \bar{c}_j
\prod^{N_f}_{f = 1}
\prod^{+ \infty}_{k = - \infty} \mathcal{D} \psi^f_k \mathcal{D} \pb^f_k
\;,
\ee
and hence
\be
\ln J = \int_x \delta \alpha \bigg\{2\mathcal{A}_3 \sum^{+\infty}_{n=-\infty}{}
\!\!\! ' \,\,\, (-1)^n - 4 \mathcal{A}_2 \sum^{+\infty}_{n=-\infty}{}
\!\!\! ' \,\,\, (-1)^n - 2 N_f \mathcal{A}_1 \sum^{+\infty}_{n=-\infty}{}
\!\!\! ' \,\,\, (-1)^n \bigg\} \; .
\ee
If the series $a_0 + a_1 + \ldots$ is Ces\`aro, Abel and Euler
summable to $s$ then $a_1 + a_2 + \ldots$ is even Ces\`aro, Abel
and Euler summable to $s - a_0$ \cite{hardy}. Therefore, from the
eq. (\ref{hardys}) follows $\sum^{+\infty}_{n=-\infty}{}
\!\!\! \!\! ' \,\,\,\,\, (-1)^n = - 1$ providing the same value
(\ref{logj}).

As in section \ref{secba}, the anomalous part of the
effective action gives the variation of the gauge coupling
constant with the scale. In fact, at low energy,{\sacolsep
\beaa
\widetilde{S}_{\mathrm{eff}}[\mathcal{J}(A), A,
\Psi\vert_{\Phi^{\textrm{\tiny{PV}}}=0}; M, M_0,\l_0] & \simeq
& \widetilde{S}_{\mathrm{eff}}[\mathcal{J}(A), A,
\Psi\vert_{\Phi^{\textrm{\tiny{PV}}}=0}; M_0, M_0,\l_0]
\nonumber \\[3mm] && + O(1/M,1/M_0) \;,
\eeaa}
which yields{\sacolsep
\beaa
S_{\mathrm{eff}}[\mathcal{J}(A), A,
\Psi\vert_{\Phi^{\textrm{\tiny{PV}}}=0}; M, M_0,\l_0] & \simeq &
\frac{1}{4}\left\{\frac{1}{g^2_0} + \left[ \frac{11}{24\pi^2} t_2(A) -
\frac{N_f}{6\pi^2} t_2(N)\right] \ln \frac{M}{M_0}
\right\} F^2_{\mu\nu} \nonumber \\[3mm] && + \cdots \; .
\eeaa}
$+ \cdots$ are terms that do not change under the considered
RG flow. Then, we obtain
\be
\frac{1}{g^2(M)} = \frac{1}{g^2_0} + \left[ \frac{11}{24\pi^2} t_2(A)
- \frac{N_f}{6\pi^2} t_2(N)\right] \ln \frac{M}{M_0} \; ,
\ee
and hence the well known result of the one-loop $\beta$-function of
QCD
\be
\beta (g) = g^3 \left[\frac{N_f}{12\pi^2} t_2(N) - \frac{11}{48\pi^2} t_2(A)\right]
\; .
\ee

\section{Conclusions}

Working with the path integral representation of the gauge
invariant effective action $\widetilde{\Gamma}[0,A]$
regularized according to the Slavnov regularization, we have
given a simple non-diagrammatic RG evaluation of the
one-loop $\beta$-function in QCD. A significant aspect of
the calculation is its compatibility with the gauge
invariance. In two respects this classical symmetry is lost
in the process of quantization: the regulator may violate
the symmetry and the gauge fixing hides the underlying gauge
invariance of the theory. In this paper we have shown the
advantages of maintaining a manifest background gauge
invariance by using a regulator that even regularizes the
divergences in a gauge invariant manner.

Another non-diagrammatic one-loop calculation has been
worked out by Fujikawa in ref. \cite{fujikawa3}, which is
based on a relation between the Weyl anomaly and the
$\beta$-function. However, our calculation being based on
the RG method, it appears to be easier to understand the
anomalous origin of the one-loop $\beta$-function in terms
of scaling of effective Lagrangians.

The method is related to one-loop but could be extended to
more then one-loop if we managed to apply the regularization
to every loop and knew a way to regulate the Jacobian like
the theory. It could also be applied in its simpler form to
get the exact $\beta$-functions of supersymmetric gauge
theories if we were able to give the supersymmetric
extension of the regularization. In fact, according to the
non-renormalization theorem, the irrelevant operators in the
Jacobian, which are $D$-terms, can be set to zero with no
change in the relevant coupling appearing in the $F$-term of
the Jacobian. Furthermore, there are suggestions that a
supersymmetric as well as gauge invariant regularization
exists \cite{slavnov3}. In particular, from West's paper
there appears to be a close relation to the regularization
scheme adopted in this paper because of the preservation of
the background and quantum gauge symmetry in addition to the
supersymmetry. Therefore, following our method, the Jacobian
could be regularized by hand as in Fujikawa's approach to
achieve the exact one-loop running of the holomorphic
coupling. This is an important point for the following
reasons. The regularization scheme mentioned above could be
used as an alternative to the Arkani-Hamed and Murayama
regularization \cite{murayama} when the proof of finiteness
appears to be scarce. Unlike the latter regularization, the
former is not limited to specific supersymmetric models.

\section*{Acknowledgments} The author would like to thank K. Yoshida, who
suggested using the PV regularization in QED and encouraged
extending the analysis to the non-Abelian case. The author
also wishes to thank S. Arnone for useful discussions.

\appendix

\section{'t Hooft's equation}
\label{appx1}

The Legendre transformation of
$\widetilde{W}[\mathcal{J},\Phi_{\tinyb}]$, assuming that
$\widetilde{\Phi}_i = \delta \widetilde{W}/\delta \mathcal{J}_i$
yields an implicit functional equation $\widetilde{\Phi}_i
= \widetilde{\Phi}_i[\mathcal{J},\Phi_{\tinyb}]$ uniquely
solvable with respect to $\mathcal{J}$, is defined as
follows:
\be
\widetilde{\Gamma}\big[\widetilde{\Phi},\Phi_{\tinyb}\big] =
\widetilde{W}\big[\mathcal{J}\big[\widetilde{\Phi},\Phi_{\tinyb}\big],
\Phi_{\tinyb}\big] - \int_x \mathcal{J}_i
\big[\widetilde{\Phi},\Phi_{\tinyb}\big]\widetilde{\Phi}_i
\;.
\ee
In this appendix we shall show that $\widetilde{\Gamma}[0,
\Phi_{\tinyb}] = \widetilde{W}[\mathcal{J}\big[\Phi_{\tinyb}],
\Phi_{\tinyb}]$ being $\mathcal{J}\big[\Phi_{\tinyb}]$ a
functional that satisfies the 't Hooft equation (\ref{eqthooft}).

The change of variables $\Phi \to \Phi - \Phi_{\tinyb}$ in the
functional integral of eq. (\ref{egfna}) results in
\be \label{relwwt}
\widetilde{W}[\mathcal{J},\Phi_{\tinyb}] = W[\mathcal{J}] - \int_x
\mathcal{J}_i\Phi^{\tinyb}_i \; ,
\ee
with $W[\mathcal{J}]$ the conventional generating functional of
connected Green's functions evaluated using the gauge fixing
surface $G^a = \partial_{\mu}(Q-A)^a_{\mu} + f^{abc} A^a_{\mu}
Q^a_{\nu}$. From the eq. (\ref{relwwt}) the identities
$\widetilde{\Phi}_i = \ovp_i - \Phi^{\tinyb}_i$ are obtained, which
show that the conditions $\widetilde{\Phi}_i=0$ are equivalent to
$\ovp_i = \Phi^{\tinyb}_i$, where $\ovp_i = \delta W/\delta
\mathcal{J}_i$. Thus, differentiating $W[\mathcal{J}]$ with
respect to $\mathcal{J}$, we must take into account the dependence
on $\mathcal{J}$ that is due to the dependence of the gauge fixing
term on the background gauge field:
\be \label{printeq}
\frac{d W}{d\mathcal{J}_i(x)} = \frac{\delta W}{\delta \mathcal{J}_i(x)}
+ \int_y \frac{\delta A^b_{\nu}(y)}{\delta \mathcal{J}_i(x)}
\frac{\delta W}{\delta A^b_{\nu}(y)} = \Phi^{\tinyb}_i(x) \; .
\ee
Note that we have distinguished a total from a partial functional
derivative with the notations $d/d\mathcal{J}$ and
$\delta/\delta\mathcal{J}$ respectively. The conditions $\ovp_i
= \Phi^{\tinyb}_i$ also give to $\mathcal{J}_i$ a dependence on
$\Phi_{\tinyb}$. Then, considering that the only explicit
$\Phi_{\tinyb}$-dependence of $W$ is on background gauge
fields, we obtain, by making use of eqs. (\ref{printeq}),
\be \label{eqdwrb}
\frac{d W}{d\Phi^{\tinyb}_i(x)} = \delta_{i1}\frac{\delta W}
{\delta A^a_{\mu}(x)} + \int_y \frac{\delta
\mathcal{J}_j(y)}{\delta \Phi^{\tinyb}_i(x)}
\frac{\delta W}{\delta \mathcal{J}_j(y)} = \int_y \frac{\delta
\mathcal{J}_j(y)}{\delta \Phi^{\tinyb}_i(x)} \Phi^{\tinyb}_j(y) \; .
\ee
Finally, we get the 't Hooft equation using eqs.
(\ref{relwwt}), (\ref{eqdwrb}) and the fermionic number.

As 't Hooft suggests \cite{hooft}, there is no need to compute
$\mathcal{J}_i[\Phi_{\tinyb}]$. Nevertheless, the class of
solutions may be restricted by the condition that the sources
$\mathcal{J}_i[
\Phi_{\tinyb}]$ transform like (\ref{gtrs}) when the background fields
$\Phi^{\tinyb}_i$ undergo the transformations (\ref{gtra}). Then
$\widetilde{W}[\mathcal{J} [\Phi_{\tinyb}],
\Phi_{\tinyb}]$ becomes a gauge invariant functional of $\Phi_{\tinyb}$ and hence
{\sacolsep
\beaa
0 & = & - \int_x \delta \Phi^{\tinyb}_i \frac{d
\widetilde{W}}{d\Phi^{\tinyb}_i}
= (-1)^{\delta_i}\int_x \delta \Phi^{\tinyb}_i \mathcal{J}_i \nonumber
\\[3mm] &=&  \int_x \big\{(D_ {\mu}\omega)^a j^a_{\mu} + i \omega^a \cb^f T^a
\psi^f_{\tinyb} - i \omega^a \pb^f_{\tinyb} T^a \chi^f + f^{abc} (\etab^a
c^b_{\tinyb} \omega^c + \bar{c}^b_{\tinyb} \omega^c \eta^a)\big\}
\nonumber \\[3mm] &=& \int_x \omega^a \big\{- (D_ {\mu} j_{\mu} )^a + i
\cb^f T^a \psi^f_{\tinyb} - i \pb^f_{\tinyb} T^a \chi^f + f^{abc} (\etab^b
c^c_{\tinyb} + \bar{c}^c_{\tinyb} \eta^b)\big\} \; ,
\eeaa}
from which we obtain
\be \label{eqthooft2}
(D_ {\mu} j_{\mu} )^a = i (\cb^f T^a \psi^f_{\tinyb} -
\pb^f_{\tinyb} T^a \chi^f) + f^{abc} (\etab^b c^c_{\tinyb} +
\bar{c}^c_{\tinyb} \eta^b) \; .
\ee

\section{Anomalous Jacobians under rescaling transformations}
\label{appx2}

Following the Fujikawa approach to the anomaly
\cite{fujikawa1, fujikawa2}, we look for the operators
appearing in the equations of motion. They can be inferred
from the quadratic part in quantum variables of the
non-regularized action in the Feynman gauge $\alpha =
1$:{\sacolsep
\beaa \label{qpqv}
&& S_{\mathrm{YM}}(A+Q) - \int_x \pb^f \left[i\dslash(A+Q) -
m^f_0\right]\psi^f - \int_x \bar{c} D^2(A+Q) c + \frac{1}{2
g^2_0}
\int_x (D_{\mu}Q_{\mu})^2 \nonumber \\[3mm] && = - \frac{1}{2 g^2_0}
\int_x Q_{\mu} (D^2 \delta_{\mu\nu} - 2i F_{\mu\nu}) Q_{\nu} - \int_x
\pb^f (i\dslash - m^f_0)\psi^f - \int_x \bar{c} D^2 c +
\cdots \; ,
\eeaa}
where $+ \cdots$ are terms which we are not interested in.
Then, under a rescaling transformation like the one in
(\ref{resctrans}){\sacolsep
\beaa
\mathcal{D}\psi \mathcal{D}\pb  \longrightarrow \mathcal{D}\psi' \mathcal{D}\pb'
& = & \dpdpb \exp \pm 2 \int_x \alpha \sum_n
\varphi^{\dag}_n \varphi_n \nonumber \\[3mm] & \equiv & \dpdpb \exp \pm 2 \int_x
\alpha \mathcal{A}_1 \; ,
\eeaa}
with the plus or minus sign when $\psi$ is a bosonic or
fermionic spinor field. In the last equation $\varphi_n$ is
a complete and orthonormal set of eigenfunctions of the
Hermitian operator $\dslash$. Therefore, the function
$\mathcal{A}_1$ is divergent. It can be regularized in a
gauge invariant manner by smoothly cutting off the
contribution of the large eigenvalues and changing the basis
vectors $\varphi_n$ for the plane wave basis as in refs.
\cite{fujikawa1, fujikawa2}: {\sacolsep
\beaa
\mathcal{A}_1 & = & \lim_{M \to \infty} M^4 \int_q \Trs f\left(q^2
- 2i\frac{(q \cdot D)}{M} + \frac{\dslash^2}{M^2}\right)
\nonumber \\[3mm] & = & \lim_{t \to 0} t^{-4} \int_q
\Trs f\left(q^2 - 2i t (q \cdot D) + t^2 \dslash^2\right)
\nonumber \\[3mm] & \equiv & \lim_{t \to 0} t^{-4} \int_q \Trs F(t) \; ,
\eeaa
where $\int_q \doteq \int d^4 q/(2\pi)^4$. The function $f(s)$ must
drop smoothly from $1$ to $0$ as $s$ goes from $0$ to $\infty$ and
$s f'(s) = 0$ at $s=0$ and $s=\infty$. Developing the matrix
function $F(t)$ in power of $t=1/M$ around $t=0$, we obtain
\be \label{anint}
\mathcal{A}_1 = \lim_{t \to 0} \sum^{4}_{n=0} \frac{1}{n!} t^{n-4}
\int_q \Trs F^{(n)}(0) + \lim_{t \to 0} \sum^{\infty}_{n=5} \frac{1}{n!} t^{n-4}
\int_q \Trs F^{(n)}(0) \; .
\ee

The second term on the right-hand side is the contribution
of the irrelevant operators, which is suppressed by negative
powers of $t$. It is zero at the one-loop level since the
regulator independent part in the first term contributes
with the correct coefficient to the one-loop
$\beta$-function, as shown in sections \ref{secba} and
\ref{secbna}. However, as pointed out in ref.
\cite{murayama}, the irrelevant operators in the Jacobian
should yield higher loop effects. In fact, according to the
RG point of view, there is an infinite number of bare
Lagrangians with the same relevant couplings and the same
low energy physics, one of which does not have irrelevant
couplings. If the Jacobian were regularized like the theory,
the operation of setting the irrelevant operators to zero
would modify the relevant coupling in the first term of eq.
(\ref{anint}) probably providing the higher order
corrections to the $\beta$-function \cite{murayama}.
Therefore, our method is related to one-loop.

Thus, being $s(t) = q^2 - 2i t (q \cdot D) + t^2
\dslash^2$ a diagonalizable matrix, the conventional rules
of derivation can be used under the trace. Then, at the one-loop
level{\sacolsep
\beaa
&& \lim_{M \to \infty} M^4 \Trs f\left(q^2 - 2i\frac{(q
\cdot D)}{M} + \frac{\dslash^2}{M^2}\right) = \lim_{M \to
\infty} \Big\{ M^4 \Trs f(q^2) - 2iM^3 f'(q^2) \Tr (q \cdot D)
\nonumber \\[3mm] && - M^2 \left[2f''(q^2) \Tr (q \cdot D)^2 - f'(q^2) \Tr
\dslash^2\right] \nonumber \\[3mm] && + M \left[\frac{4i}{3} f^{(3)}(q^2)
\Tr (q \cdot D)^3 - 2i f''(q^2) \Tr \dslash^2 (q \cdot D)\right]
\nonumber \\[3mm] && + \frac{2}{3} f^{(4)}(q^2) \Tr (q \cdot D)^4
- 2 f^{(3)}(q^2) \Tr \dslash^2 (q \cdot D)^2 +
\frac{1}{2} f''(q^2) \Tr \dslash^4 \Big\} \; .
\eeaa}
Finally, by making use of the integrals
\be
\begin{array}{l}
{\ds \int_q f(q^2) q_{\mu_1} \cdots q_{\mu_n} = 0 \qquad
\mathrm{for} \;\: \mathrm{odd} \;\: n,} \\[0.6cm] {\ds \int_q f(q^2) q_{\mu}
q_{\nu} = \frac{1}{4} \delta_{\mu\nu} \int_q f(q^2) q^2 \; ,}
\\[0.6cm] {\ds \int_q f(q^2) q_{\mu} q_{\nu} q_{\rho} q_{\sigma}=
\frac{1}{24}\left(\delta_{\mu\nu} \delta_{\rho\sigma} + \delta_{\mu\rho} \delta_{\nu\sigma} +
\delta_{\mu\sigma} \delta_{\nu\rho}\right) \int_q f(q^2) q^4 \; ,}
\end{array}
\ee
and the property of $f(s)$, we obtain
\be
\mathcal{A}_1 = \lim_{M \to \infty} M^4 \Tr \int_q f(q^2) +
\frac{1}{24\pi^2} \Trg F^2_{\mu\nu} \; ,
\ee
where $\Trg$ means a trace only on the gauge group indices. For our
purpose, the first term on the right-hand side can be left out as
field independent. The same will be done in the following
calculation of $\mathcal{A}_2$ and $\mathcal{A}_3$.

The anomalous Jacobians under rescaling transformation of scalar
and vector fields are evaluated by the same procedure:{\sacolsep
\beaa
\!\!\!\!\!\!\!\!\! && \mathcal{D}c \mathcal{D}\bar{c} \longrightarrow \mathcal{D}c'
\mathcal{D}\bar{c}' = \mathcal{D}c \mathcal{D}\bar{c} \; \exp \pm 2 \int_x \alpha
\sum_n \vartheta^{\dag}_n \vartheta_n \equiv \mathcal{D}c \mathcal{D}\bar{c} \;
\exp \pm 2 \int_x  \alpha \mathcal{A}_2 \; , \\[2mm] \!\!\!\!\!\!\!\!\! &&
\mathcal{D}Q \longrightarrow \mathcal{D}Q' = \mathcal{D}Q \; \exp
\pm \int_x \alpha \sum_n \varrho^{\dag}_n \varrho_n \equiv
\mathcal{D}Q \; \exp \pm \int_x  \alpha \mathcal{A}_3 \; ,
\eeaa}
where, according to the eq. (\ref{qpqv}), $\vartheta_n$ and
$\varrho_n$ are complete and orthonormal sets of
eigenfunctions of the Hermitian operators $D^2$ and $D^2
\delta_{\mu\nu} - 2i F_{\mu\nu}$ respectively. The sign
follows the same previous rules. $D_{\mu}$ is an
anti-Hermitian operator with respect to the inner product
$(c,c) = \int_x c^*_a c^a$ and therefore $D^2$ is positive
semi-definite. Then, suppressing the contribution of large
eigenvalues as above, at the one-loop level we get
\be
\mathcal{A}_2 = \lim_{M \to \infty} M^4 \int_q \Trs f\left(q^2
- 2i\frac{(q \cdot D)}{M} - \frac{D^2}{M^2}\right) = - \frac{1}{192\pi^2}
\Trg F^2_{\mu\nu} \; .
\ee
Repeating the same procedure for the calculation of
$\mathcal{A}_3$, we obtain
\be
\mathcal{A}_3 = \frac{5}{48\pi^2} \Trg F^2_{\mu\nu} \; .
\ee

These results have also been worked out in refs.
\cite{fujikawa2, fujikawa3} as flat space-time limit of the
Weyl anomaly.

\end{document}